\documentstyle[spie]{article} 

\title {Mesosphere Sodium Column Density and the Sodium  Laser Guide 
Star Brightness} 

\author{Jian Ge\supit{a}, J.R.P. Angel\supit{a}, B.D. Jacobsen\supit{a}, 
T.  Roberts\supit{a}, T. Martinez\supit{a}, W. Livingston\supit{b},\\
 B. McLeod\supit{c}, M. Lloyd-Hart\supit{a}, 
P. McGuire\supit{a}, R. Noyes\supit{c}
\skiplinehalf 
\supit{a}Steward   Observatory, The University of Arizona,
Tucson, AZ \hspace{0.3em}85721, \hspace{0.4em}USA 
\skiplinehalf 
\supit{b}National Solar  Observatory, Tucson, AZ \hspace{0.3em} 85726, 
\hspace{0.4em}USA
\skiplinehalf 
\supit{c}Center for Astrophysics, MS  20,  60 Garden  Street, Cambridge,  MA \hspace{0.3em} 02138, \hspace{0.4em}USA }

\authorinfo{Other author information: (Send correspondence to Jian Ge)\\J. G.: 
Email: jge@as.arizona.edu;  WWW: http://qso.as.arizona.edu/$^\sim$jge;
 Telephone: 520-621-6535; Fax: 520-621-1532}

\begin{document}
\maketitle 

\begin{abstract}
The first time simultaneous measurements of sodium column density and the 
absolute flux 
from a sodium laser guide star, created by a monochromatic 3 W cw laser, tuned 
to the peak of the sodium D$_2$ hyperfine structure, were conducted at the MMT 
and CFA 60 inch telescope in 1997. The results show that linearly and 
circularly polarized laser returns are 
proportional to the simultaneous sodium column density. Moreover,   
circularly polarized laser provides $\sim$ 30\% increase in fluorescent
return over linearly  polarized laser. A laser guide star with R = 10.3
mag. or absolute flux  of 8.4$\times 10^5$ photons s$^{-1}$ m$^{-2}$, could be 
formed from a 1  watt projected circularly polarized sodium laser beam when 
sodium layer abundance N(Na) = 3.7$\times  10^9$ cm$^{-2}$.  
Together with the distributed column density measurements (e.g. seasonal  and
diurnal variations), we can project laser power requirements for any 
specified guide star brightness.

The mesosphere sodium column density variation  was measured above Tucson
sky throughout the year, through sodium absorption line  measurements in 
stellar and solar spectra. Previous measurements, e.g. Papen 
et al, 1996, have not been made at this latitude (32 degrees).  Further, 
our absorption method is more direct and may be more accurate than the 
lidar methods normally used. The  seasonal variation  amplitude
is smaller than that at  higher latitudes. While the annual  mean sodium column 
density tends to be lower than at  higher latitudes. Diurnal sodium column
density  tends to vary by as much as a factor of  two  within an hour.

\end{abstract}



\section{INTRODUCTION}

Adaptive  optics systems for atmospheric-turbulence compensation
 require a bright point reference source   of  light for measuring and  
correcting wave-front distortions.  The bright source must be  within a small 
field of  view of  the astronomical sources of interest.  
While some sources are bright enough to provide the  required 
information themselves, most astronomical sources  of interest are too
faint. For  this  reason, general  astronomical use   of adaptive optics
requires  a  laser  beacon to provide  the wavefront  information. An
artificial  laser  beacon  can be created by backscattered light from a  
ground-based laser beam  pointed at the  scientific  source.  For example,
an artificial guide  star  is created by focusing a sodium laser  beam tuned to
the sodium D$_2$ line  at  589 nm  wavelength on the 
mesosphere sodium layer  at  about 90 km altitude. 
 Several groups  in  the world have  begun to explore  the use of sodium 
laser guide star   technique
for  astronomical adaptive optics  (e.g. Jacobsen et al. 1993;
 Max  et  al. 1994; Jelonek et al. 1994;
Avicola et al. 1994). In the Fall 1996, our group has 
successfully  closed  the laser  guide star adaptive optics tip-tilt 
system loop and make improvements  in  the image  quality (Lloyd-Hart et al. 
1997). In the early 1997,  Livermore laser guide star  group  has also 
successfully  closed their high order AO loop  and make  significant image 
corrections (Olivier et al. 1997).   These experiments demonstrate the 
possibility  of  sodium laser  guide  star technique in adaptive optics 
applications. However, for the sodium laser guide star adaptive optics 
technique, there is as yet no consensus on the optimum laser and power.
  The choice depends not only on the laser power, but on the inherently different pulse formats and frequency purity of different 
laser types.  Theoretical calculations predict differences in return for the 
same power and column density up to a factor 5 (Milonni \& Telle, 1996, private
communication), according to pulse format, but 
these have yet to be confirmed experimentally.  Thus, experiments that measure 
column density and details of the excitation and scattering properties of sodium atoms in the sodium layer are very important to refine the design parameters of 
the laser and assess the power requirement.

Simultaneous local observations of the sodium column density and the
laser return are necessary to obtain the fundamental
relationship between the magnitude of the laser
guide star and sodium abundance.  Previous
studies have shown that the column density of the sodium layer varies
with time, including long term seasonal variations and short term
variations (e.g. Papen 1996), and also varies with latitude (Hunten 1966; 
Magie  et al. 1978; Papen et al. 1996).  A
simultaneous measurement of column density and guide star brightness
has not been done before.  
This together  with the measurements
of sodium layer abundance variation will provide a reference
 for the design of our
future sodium laser for the MMT 6.5 m AO system as well as  other sodium laser guide  star systems in the world.

In  this  paper we will briefly report on seasonal  sodium layer column  
density variations over Tucson (Ge, Angel, \& Livingston 1997, in preparation),
simultaneous sodium laser return and
mesosphere sodium abundance   measurements.
 We will compare the  new results  with previous results  by
other groups  and  finally draw  some initial conclusions. 

\section{Observations  and Reductions}

\subsection{Seasonal Mesosphere Sodium Observations and Reductions}

The observations of the seasonal mesosphere sodium variation over  Tucson
were conducted  with the National  Solar Observatory McMath solar telescope 
on Kitt Peak.  The high resolution 13.5 m main spectrometer
with the 632 g/mm grating and predisperser was used  to provide 
resolution of about 2.6$\times 10^6$ (theoretical value). The receiver is 
a photomultiplier  with no readout noise, no bias and no cosmic ray. 
Solar sodium D$_2$
absorption spectra are one-dimensional and  were obtained in 1 minute 
integration.  

The main method for the data reduction is to fit the generated Voigt profile
to the bottom  of the solar D2 line.
 The main object of the  computer program
was  to compute a set of optimum parameters that could generate the best 
least-square fit of the model  to the data (excluding the mesosphere  D$_2$ 
absorption region).
We then extracted the solar absorption by removing the
fitted profile to obtain the residual spectrum which contains the tiny
absorption lines from the sodium layer. The typical solar spectrum near the 
bottom of  the D$_2$ sodium line and residual telluric D$_2$ absorption 
spectra are shown in Figure 1.

	\begin{figure}	
	\vspace{5.0cm}	
\includegraphics{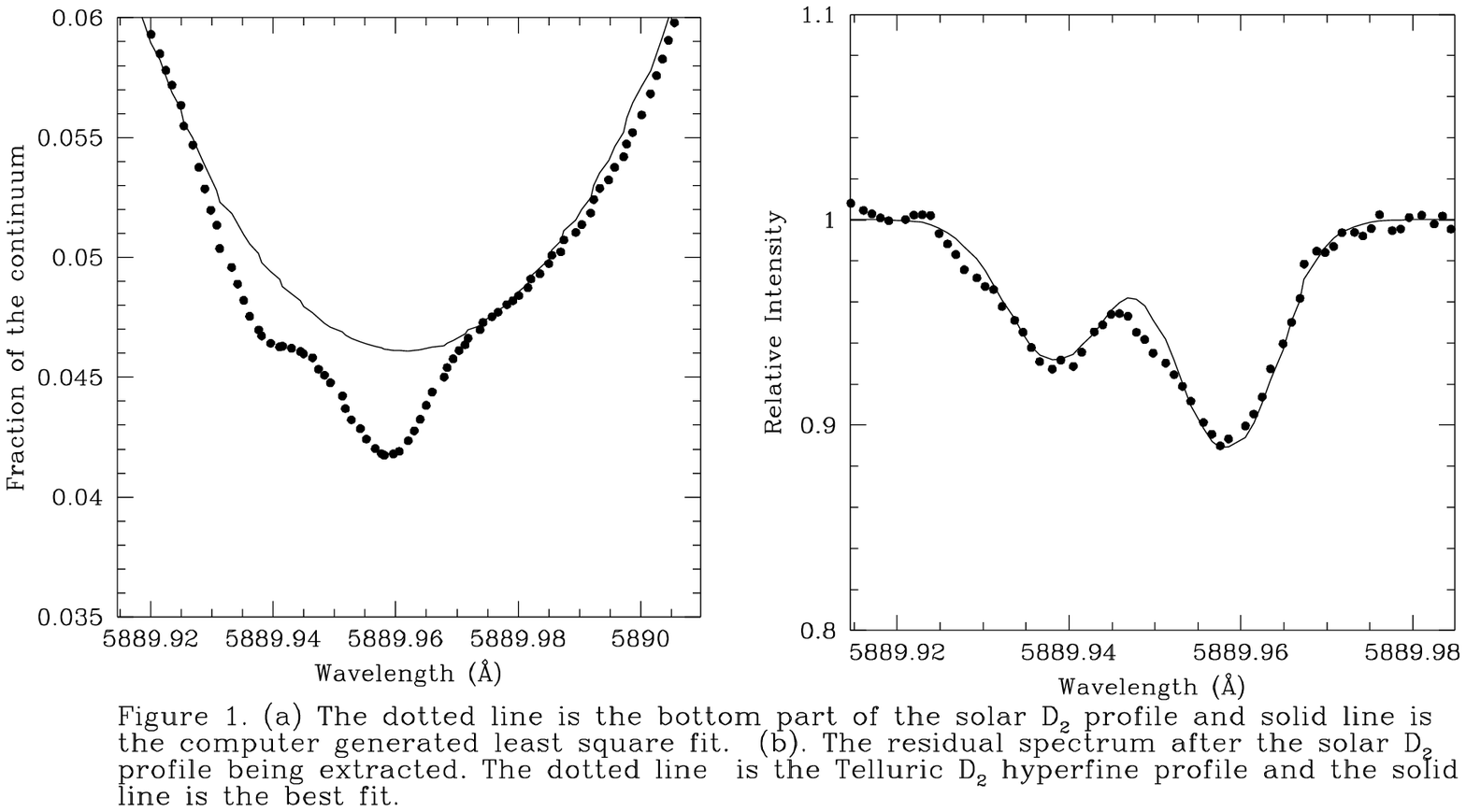}
	\end{figure}

\subsection{Simultaneous  Sodium Laser Return and  Column Density Measurements}

The simultaneous measurements of  mesosphere  sodium abundance and  laser 
return were made in the nights of March 23rd and May 20th, 1997.
 Both  nights were photometric. The sodium 
 return experiments were conducted at the MMT on Mt. Hopkins. The telescope was
pointed to the direction of standard  stars, which were close to the stars  
for observing telluric sodium absorption.  At the same time, telluric sodium
absorption spectra were recorded  by the Advanced Fiber Optic  Echelle  
spectrograph (AFOE) at the CFA 60$''$ telescope about 1 km away from the 
MMT on the same mountain. Therefore, similar sodium layer
patches were observed by  both telescopes.

The laser we applied is a continuous-wave dye laser pumped by 30 watts argon 
ion laser. A traveling-wave ring cavity design for  the dye laser was used
 to  provide a single longitudinal mode at high powers. 
 Tuning of the ring-cavity laser was made by 
adding several frequency selective elements such as a three-plate birefringent 
filter and  two etalons to the cavity.
The detailed  design   and performance are described by Roberts  et al.
(1997). The laser system produced a stable sodium laser output  power  of
about 2.4 Watts,  tuned to the peak of D$_2$ hyperfine structure  
during the return experiments.  The full width at half maximum (FWHM) of the 
laser line profile is less than 10 MHz. However, vibrations in the laser system,
especially in the dye jet, could produce  frequency jitter of about 0.7  GHz 
(peak-to-valley). 
 A polarization  compensator made of MgF$_2$ crystal 
 has  been introduced in the  sodium laser  beam
to change the laser polarization during the  observations. After
the sodium laser passed through the whole laser beam  projector system 
(Jacobsen \& Angel 1997),  power projected
to the sodium layer was  typically 0.9W. There are 14 transmissive and 7 
reflective  surfaces between the Dye  laser output and   the laser 
beam projector output, which provides about total 45\% transmission. 
The R filter used for the photometric  measurements transmits 79\% photons 
at the D$_2$ wavelength. Therefore, the total throughput of the laser
 projection is about 36\%, which is consistent with the sodium laser  power 
measurement results. 
 The laser  star and standard star images  were  recorded  by a thermally 
cooled Axiom CCD camera mounted  on the telescope. The data were reduced in  the
standard way with the IRAF package. The photometry of laser guide stars
was measured  by using the IRAF package PHOT. The estimated error of the 
photometry is 2.5\% based on the standard star flux measurements.

The spectrograph we used for measuring the telluric sodium abundance provides
spectral resolution  of R  $\sim$ 50,000, which cannot provide enough 
resolution to  separate D$_2$ line from nearby water  vapor  lines. We 
therefore use D$_1$, which has half the D$_2$ line  strength, instead.
The equivalent  width of typical telluric sodium D$_1$ absorption line is 
 less than 1 m\AA. In order
to measure the sodium abundance  better  than 15\% error,  signal  to noise
ratio (S/N) of at least  1,000 is  required. With 16  bits CCD readout 
electronics, S/N of   $\sim$ 500 can be reached in each frame. On the other
hand, once  S/N reaches about 500, CCD pixel-to-pixel  variation  will limit
S/N go any higher. To solve this problem, we  took $\sim$ 100  frames of flat  
and combined
them,  and divided the combined object frames ($\alpha$ Leo in the March
run, and $\alpha$ Aql in the May run) by the combined flat. The final S/N 
of combined object  frame can reach as  high as $\sim$ 2,000, which is good  
enough  for the accurate sodium abundance measurements. Fig. 2(a),(b) show 
typical telluric D$_1$ spectra  from both observation  runs. 

	\begin{figure}	
	\vspace{6.5cm}	
\includegraphics{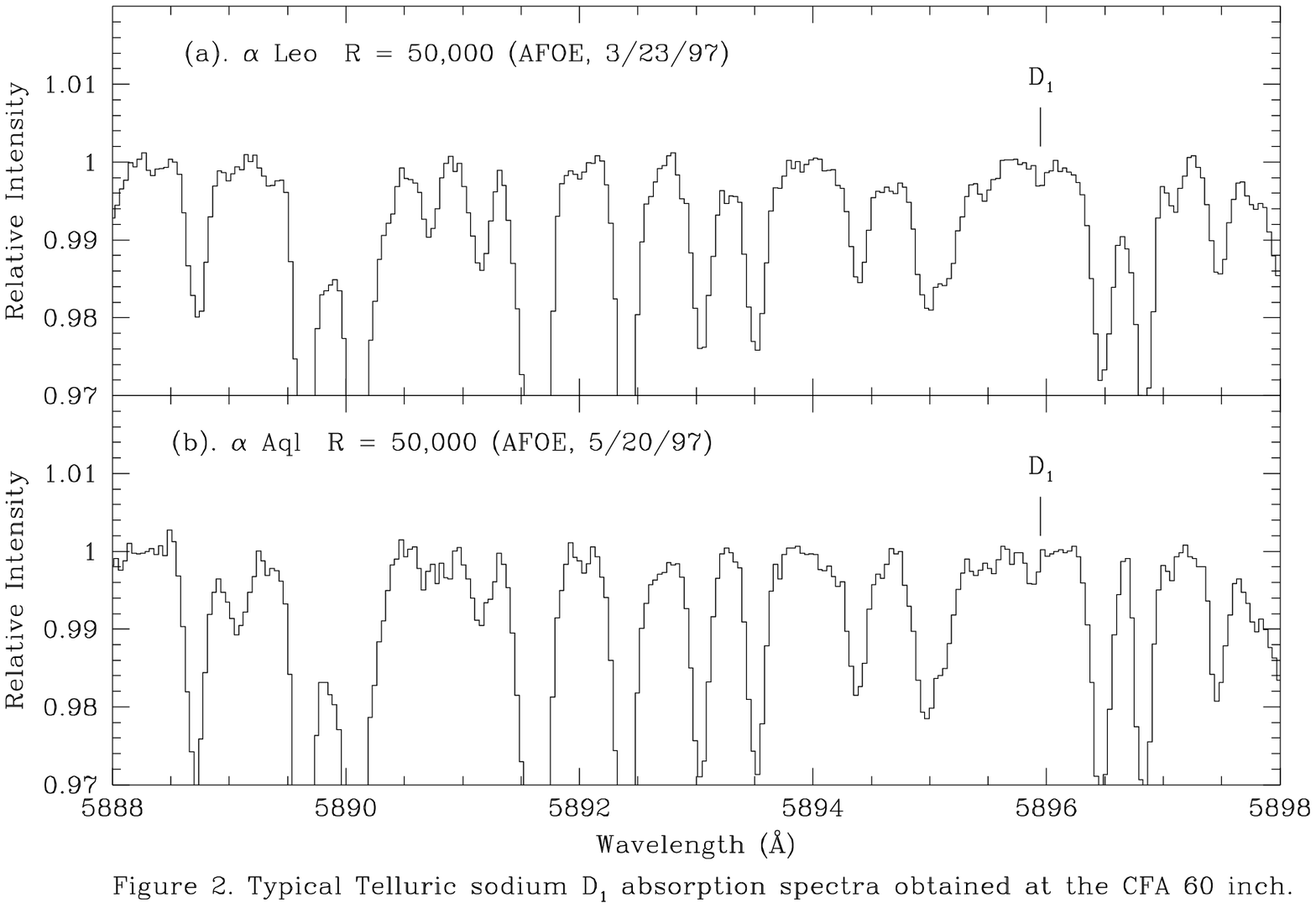}
	\end{figure}

\section{Results}

\subsection{Seasonal Sodium Abundance Results}

Because the mesosphere sodium absorption  is normally  very weak, e.g.  the
typical optical depth at the center of the absorption line is $\tau_0 \sim  
0.04 << 1$, it is appropriate  to  assume that the absorption line is in the
linear portion of the curve  of growth, where the column density of the atoms
is proportional to the equivalent width. The number of sodium atoms per square centimeter in the line of sight is then given by
\begin{equation}
N = \frac{mc^2}{\pi e^2 f}\int_0^\infty \frac{I_0(\nu) - I(\nu)}{I_0(\nu)}  d\nu = 1.130\times 10^{20}  \frac{W_\lambda}{f  \lambda^2}~ cm^{-2},
\end{equation}
where $W_\lambda$ is the  equivalent width,  and both $W_\lambda$ and $\lambda$
are in  unit of \AA, $f$ is the oscillator strength ($f = 0.3944, 0.2367$ for the 5889.9584 \AA, 5889.9386 \AA\ hyperfine
lines  of  the D$_2$, respectively; Morton 1991).  The error resulted  from
 neglecting saturation effect was
estimated to be less than about 2.5\%, 
which is within the measurement  error of about 5\% for the whole seasonal 
observation results. 

Figure 3 shows the seasonal variations of the sodium layer column density 
above Tucson between  February 1995 and  January 1996. Since we  observed 
the  Sun at different airmass, we  need to normalize the measured
column densities to the values at the zenith. The relationship between the 
airmass and the sodium layer thickness can be expressed as
\begin{equation}
sec\phi=\frac{1}{\sqrt{1-(\frac{r_e}{r_e+H})^2(1-\frac{1}{sec^2z})}}
\end{equation}
where $sec \phi$ is the thickness factor of the sodium layer,
$r_e$ = 6378.5 km is the radius of the Earth,  the height of the sodium layer
above Kitt Peak, $H$ = 90 km is assumed, and $sec~z$ is the airmass. 
Thus, the value of $sec\phi$ is 1.0  when $sec~z$ = 1.0 and it is about 6.0
 when $sec~z$ $\rightarrow \infty$.
 The column densities in Figure 3 represent the values at the zenith.  Each 
point represents  an  averaged value within 4-8 minute integration time. The
dominant error in these measurements is from the solar D$_2$ profile fitting,  which is  estimated to be about 5\%. The  scattering in the measured
values during each  day therefore  represents the real  short term  variation.
The  variation amplitude could be as large as  a factor of  two  within a  short
period of  time (on the order of  hours),  which has also  been found 
 at other latitude by Lidar technique (e.g. Papen et al. 1996). 
There is  also a strong  trend of seasonal variation  in this
figure.   Higher sodium  abundance was reached  during the winter time with  
an average of $\sim 4\times 10^9$ cm$^{-2}$, except October data points, when  
the sodium abundance is extreme large  compared  with the  values in the neighbour months. This extreme high abundance could  be caused  by sporadic  
events  during that day, caused  by  such  as 
meteor events or geomagnetic activity (Papen et al. 1996).  Generally, lower  sodium  abundance  was 
reached  during the  summer time with  an average  value of $\sim 2\times 10^9$
cm$^{-2}$.  The annual mean is 3.7$\times 10^9$ cm$^{-2}$ is lower than that 
at higher latitude (Papen et al. 1996). If we neglect the 
data points from the October  run,  the variation amplitude within a year 
period from Tucson (32$^\circ$ N) is smaller 
than that at higher latitudes (e.g. Urbana, 40$^\circ$ N. Papen et al. 1996; 
Haute-Provence, 44$^\circ$ N. Magie  et al. 1978).
These new measurements from different latitudes  confirm the latitude-related 
trend shown
 in  earlier studies using twilight emission line  technique (Hunten  1966).
  Detailed  analysis on the seasonal, daily, nightly 
and hourly variation   of sodium abundance above
Kitt Peak can be seen in  our future  paper (Ge, Angel \& Livingston 1997). 

	\begin{figure}	
	\vspace{4.5cm}	
\includegraphics{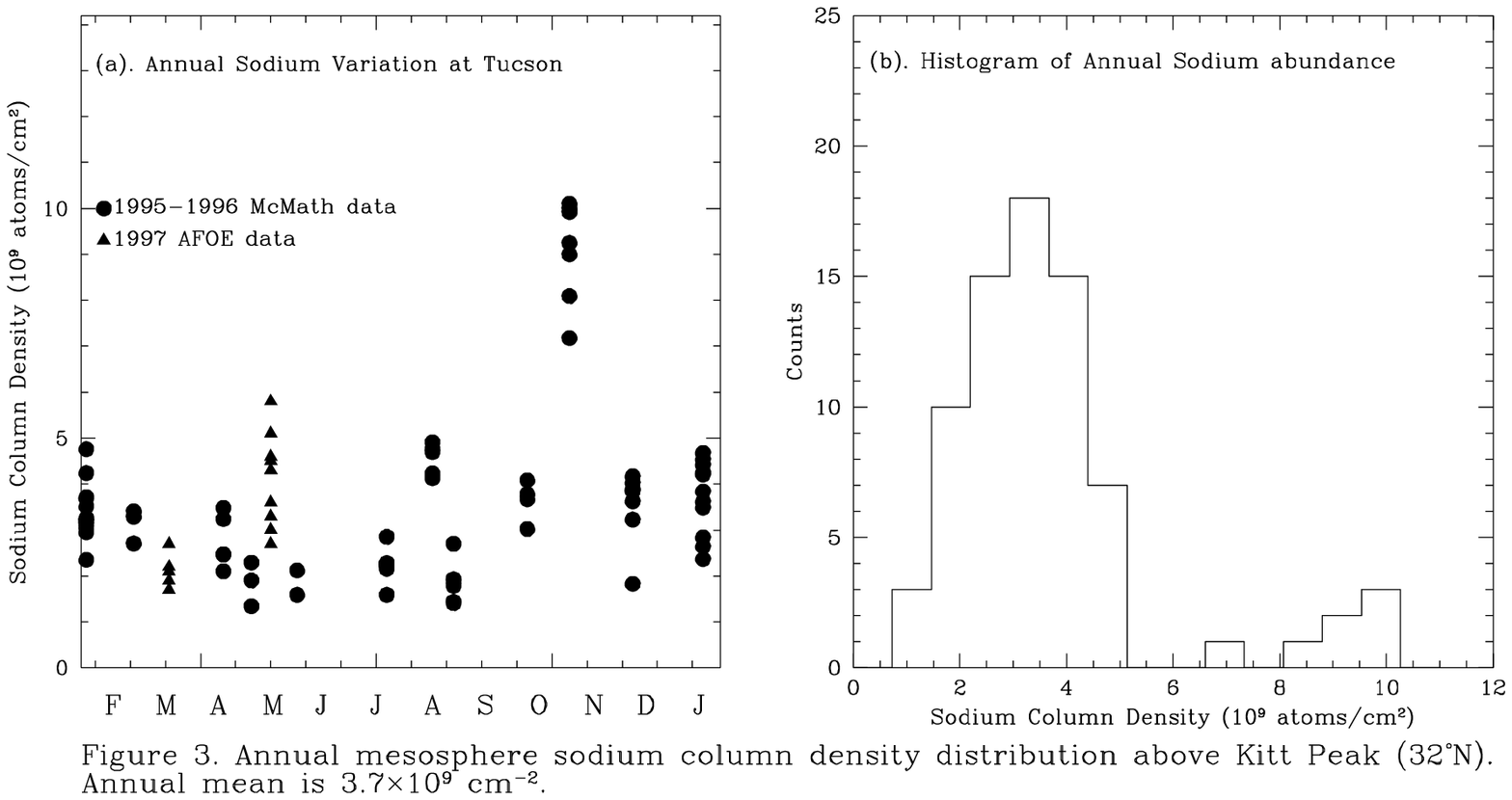}
	\end{figure}

\subsection{Simultaneous Sodium  Return and  Abundance Results}

Table 1   and 2 show the measurements of  simultaneous absolute return flux 
from linearly and circularly polarized sodium laser beams and  sodium
column density from the March and May runs, 1997, which also include
 corresponding R magnitude. Fig. 4(a)(b)   show
the relationship between the absolute laser  guide  star flux and sodium 
abundance at the  zenith.   Both backscatters from circular and linear
polarization   are   proportional to sodium column density. The absolute
return flux range is consistent with that from
our  previous return measurements  though
no  simultaneous sodium abundance being measured at that time
 (Jacobsen et al. 1993). Further, the
circular  polarization   provides  $\sim$ 30\% increase in fluorescent
return over linear polarization. Based on these results,  a laser guide star 
with R = 10.3 mag. or absolute flux  of 8.4$\times 10^5$ photons s$^{-1}$ 
m$^{-2}$, could be formed from a 1  watt projected circularly polarized sodium 
cw laser beam when sodium layer abundance N(Na) = 3.7$\times  10^9$ cm$^{-2}$,
which is the mean value  of annual sodium abundance in the sodium layer  
 at the latitude of 32$^\circ$ N (Tucson).

\begin{table} [h]   
\caption{Simultaneously Linearly Polarized Laser Return and Column 
Density Measurements}
\begin{center}
\begin{tabular}{lcccc}  \hline\hline
No.& Time (MT) & Absolute Return Flux at Zenith & Sodium Column Density at Zenith& R magnitude/W\\
   &      & (10$^5$ Photons/m$^2$/s/W)&  (10$^9$ atoms/cm$^2$)& \\
\hline
1 & 1h13m, 3/23/97& 2.6 & 2.2$\pm$0.5 & 11.5\\
&&&&\\
2 &3h18m, 5/20/97& 6.6 & 3.0$\pm$0.5 & 10.5 \\
3 &3h23m, 5/20/97& 9.6& 4.3$\pm$0.7 & 10.1 \\
4 &3h33m, 5/20/97& 8.5& 4.6$\pm$0.7 & 10.3\\
5 &3h39m, 5/20/97& 8.8& 4.3$\pm$0.7 & 10.2\\[0.0ex]
\hline

\end{tabular}
\end{center}
\end{table}

\smallskip

\begin{table} [h]   
\caption{Simultaneously Circularly Polarized Laser Return and Column 
Density Measurements}
\begin{center}
\begin{tabular}{lcccc}  \hline\hline
No.& Time (MT) & Absolute Return Flux at Zenith & Sodium Column Density at Zenith& R magnitude/W\\
   &      & (10$^5$ Photons/m$^2$/s/W)&  (10$^9$ atoms/cm$^2$)& \\
\hline
1 &12h32m, 3/23/97& 3.8 & 1.7$\pm$0.5 & 11.1\\
2 &1h16m, 3/23/97& 3.7 & 2.1$\pm$0.4 & 11.2\\
3 &1h26m, 3/23/97& 4.4 & 1.9$\pm$0.5 & 11.0\\
4 &1h55m, 3/23/97& 4.5 & 2.7$\pm$0.4 & 11.0\\
&&&&\\ 
5 & 2h45m, 5/20/97& 11.3 & 4.5$\pm$0.7 & 10.0 \\
6 & 3h08m, 5/20/97& 9.0 & 3.3$\pm$0.6 &10.2  \\
7 & 4h04m, 5/20/97& 7.7 & 3.6$\pm$0.7 & 10.4\\
8 & 4h08m, 5/20/97& 5.1 & 2.7$\pm$0.5 & 10.8\\
9 & 4h14m, 5/20/97& 11.3 & 5.1$\pm$0.9 & 10.0 \\
10 & 4h18m, 5/20/97& 12.3 & 5.8$\pm$0.9 & 9.9 \\
\hline

\end{tabular}
\end{center}
\end{table}   

	\begin{figure}	
	\vspace{4.5cm}	
\includegraphics{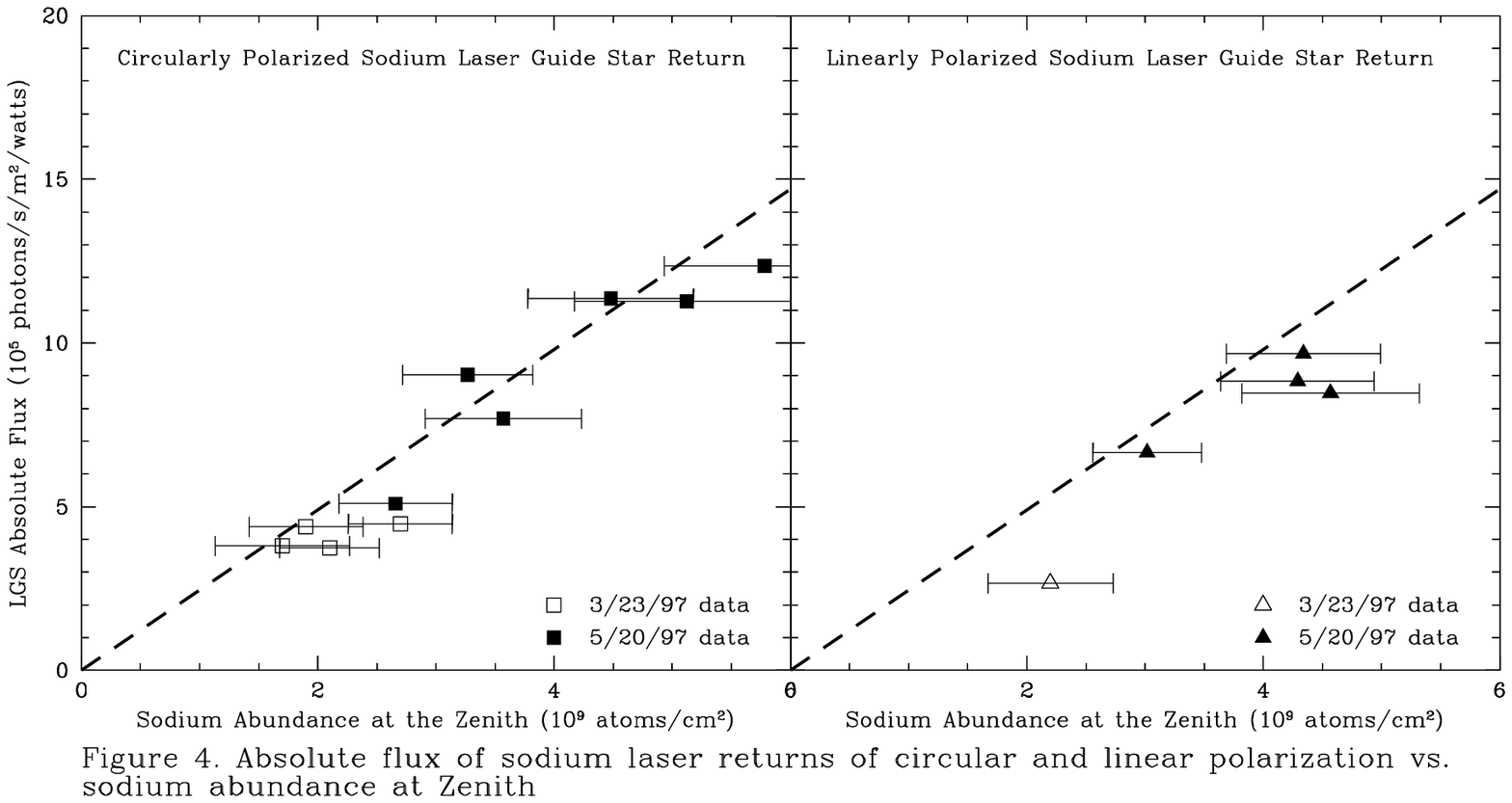}
	\end{figure} 

The backscatter  enhancement by 
circular polarization is consistent  with the theoretical prediction (Morris 
1994; Milonni, Fugate \& Telle 1997), and also qualitatively consistent with the LGS measurement report by 
Jelonek et  al. (1994). The  dashed lines in Fig. 4(a)(b) are the theoretical 
curves generated from the following  simple theoretical estimate
\begin{equation}
 N_r = N_t \epsilon \frac {A_{c}}{4\pi  H^2}  T^2 1.5,
\end{equation}
where $N_r$ is backscatter in unit of  photons s$^{-1}$, $N_t = 3.0\times 10^{18}$ photons s$^{-1}$  is the total output flux of 1 watt power, $A_c$ is the photon 
collector area, $H = 90$ km, atmosphere transmission $T\approx 0.8$ during the experiments.  $\epsilon = \sigma_P  N_{Na}$,  where $N_{Na}$ is the sodium 
column density, and peak cross
section of  D$_2$ hyperfine $\sigma_P = 8.8\times 10^{-12}$ cm$^2$ is applied
because  the line width of $<$ 10  MHz for the cw dye laser tuned to the
hyperfine peak is very narrow compared  to 1.19 GHz for the FWHM of 
D$_2$ hyperfine line  (Happer et al. 1994).
 The angular  dependent  scattering factor, 1.5, for the backward
is also used in equation (3) (Jeys  1991). 
The simple theoretical prediction matches  the circular  polarization 
measurements very well. 
However,  the results from preliminary theoretical studies 
including the Earth's magnetic field and optical pumping effects predicted a
factor of two higher return than  that we have measured (Milonni 1997,  private 
communication). The difference could be caused by the premature theoretical 
studies  or possible less measured  return flux caused  by the sodium laser frequency instability during the return experiments.
 If the laser frequency  instability is confirmed in the future
MMT sodium guide star experiments, then  we  should consider this frequency 
instability effect in our measured results. 


\section{DISCUSSION}
Annual mesosphere sodium abundance measurements show that  there is long term 
 seasonal  variation at the latitude of 32$^\circ$ N.,
  and  variation amplitude is about a factor 
of  two. Sodium  column density could also vary by a  factor  of  two with a 
short  period  of  time (hourly). Therefore the laser power dynamical 
range should be  at least designed to  match  the sodium abundance 
variation range of a factor of two in order to provide enough  return photons 
for LGS AO system to get enough image  correcting power. The preliminary 
results  from the simultaneous measurements of the sodium laser return  and 
sodium layer  abundance  suggest that a circularly polarized sodium 
cw  dye single  longitudinal mode laser with 3  watts projected power on sky 
tuned to the peak of D$_2$ hyperfine  will 
meet the LGS brightness requirement of 9.5   mag. for the future MMT 6.5 m telescope LGS AO system (Sandler et al. 1994), even when sodium abundance 
reaches the lowest point in the  seasonal variation curve (Fig. 3).


\acknowledgements
We thank  MMT staffs for great patience and  help during our FASTRAC  II run.
We thank Perry Berlind   and Jim Peters  for helping collecting  and reducing  
60 inch AFOE  raw data. We thank L. Wallace for  helpful discussions. 
We thank D. Lytle for his helpful program to convert the raw solar date  
format to standard IRAF format. 
This work has been supported by the Air Force Office
of  Scientific Research under grant number F49620-94-1-00437 and F 
49620-96-1-0366.

\vspace{0.5cm}
\centerline {\bf REFERENCES}
\noindent Avicola, K. et al. 1994, JOSA, 11, 825\\
\noindent  Happer, W., MacDonald, G.J., Max,  C.E., \& Dyson, F.J. 1994,  JOSA, 11, 263\\
\noindent Hunten, D.M.  1967, Space Science  Reviews, 6, 493\\
\noindent Jacobsen, B. et al. 1993, SPIE, 2201, 342\\
\noindent Jacobsen, B. \& Angel, R., 1997, this proceeding\\
\noindent Jelonek, M.P. et  al. 1994, JOSA, 11, 806\\
\noindent Jeys, T.H. 1991, The Lincoln  Lab. Journal, 4,  133\\
\noindent Lloyd-Hart, M. et al. 1997, ApJ Letter, submitted\\
\noindent Max, C.E. et al. 1994, JOSA, 11, 813\\
\noindent Magie, G.  et al. 1978, Planet. Space Sci. 26, 27\\
\noindent Milonni, P.W., Fugate, R. Q., \&   Telle,  J.M. 1997, in preparation\\
\noindent Morris, J.R. 1994, JOSA, 11, 832\\
\noindent Morton, D.C. 1991, ApJS, 77, 119\\
\noindent Olivier, S.S. et al. 1997, SPIE, 3126, in press\\
\noindent Roberts, T., et al. 1997, this proceeding\\
\noindent Papen, G.C., Gardner, C.S. \& Yu, J. 1996, in Adaptive 
Optics, Vol. 13, OSA Technical Digest Series (Optical Society of America, Washington DC), 96\\
\noindent Sandler, D.G., et al. 1994, JOSA, 11, 925

\end{document}